\documentclass[11pt]{article}
\usepackage{amssymb}
\def\nn{\nonumber}

\def\l{\left}
\def\r{\right}

\def\beq{\begin{equation}}
\def\eeq{\end{equation}}
\def\bea{\begin{eqnarray}}
\def\eea{\end{eqnarray}}
\def\barr{\begin{array}}
\def\earr{\end{array}}
\def\be{\begin{equation}}
\def\ee{\end{equation}}
\def\bc{\begin{center}}
\def\ec{\end{center}}

\def\Ne#1{{\cal N}=#1}
\def\De#1{{\it D}=#1}
\def\Zd{{\cal Z}_2}

\def\fso{\mathfrak{so}}
\def\fsu{\mathfrak{su}}
\def\fusp{\mathfrak{usp}}

\begin{document}
\begin{titlepage}
\begin{flushright}
CERN-PH-TH/2004-130 \\
ROMA-1383/04
\end{flushright}
\vskip 2cm
\begin{center}
{\LARGE \bf $\bf {\cal N}=6$ gauged supergravities \\ \vspace{0.3cm} from generalized dimensional reduction} 
\vskip1cm {\large Giovanni Villadoro}\\ \vspace{.5cm}
{\normalsize {\sl 
Theory Division, Physics Department, CERN, CH-1211 Geneva 23, Switzerland, \\ 
Dipartimento di Fisica, Universit\`a di Roma ``La Sapienza'', and \\ 
INFN, Sezione di Roma, P.le A. Moro 2, I-00185 Rome, Italy}} \\ \vspace{0.25cm}
{e-mail: giovanni.villadoro@roma1.infn.it}

\vskip2.0cm {\large\bf Abstract\\[10pt]} \parbox[t]{\textwidth}{{
We construct new $\Ne6$ gauged supergravities in four and five dimensions using generalized dimensional reduction.
Supersymmetry is spontaneously broken to $\Ne4,2,0$ with vanishing cosmological constant. We discuss the gaugings 
of the broken phases, the scalar geometries and the spectrum. Generalized orbifold reduction
is also considered and an $\Ne3$ no-scale model is obtained with three independent mass parameters.
}}
\end{center}
\vspace{4cm}
 {\small {PACS numbers}: 04.50.+h, 04.65.+e, 11.25.Mj, 11.30.Qc}. \hspace{0.3cm}  

\end{titlepage}
\baselineskip 6 mm
\section{Introduction}

It has recently been shown that generalized dimensional reduction \cite{ss}
from $D$ to $D-1$ dimensions \cite{fg28,vz,dual1,dual2}, as well as orientifold reduction with fluxes \cite{fluxes1,fluxes2,fluxes3},
can be used to derive new gauged supergravities with spontaneous 
supersymmetry breaking, which escaped previous classifications \cite{classif}. 
The reason why this type of gaugings had not generally been 
considered before is that dimensional reduction does not give the {\em standard}
effective theory, but a dualized one, where a subgroup of the $R$-symmetry is embedded in
a hidden sector \cite{crem}. This happens because, in the higher-dimensional theory, part of the supersymmetry
automorphism group is contained in the Lorentz group. To obtain these supergravities 
directly in lower dimensions, it is necessary to move first in the dualized theory, 
then perform the gauging. This is in fact the procedure followed in \cite{tz3,tz4+d} 
to derive these theories for $\Ne3,4$. 
Generalized reduction represents a very powerful tool to do this
systematically and it was used in \cite{fg28,vz} to get new $\Ne2,4,8$ gauged 
supergravities in diverse dimensions. In this letter we apply it 
to derive $\Ne6$ no-scale models in $\De4$ and $\De5$.

The Scherk--Schwarz twist is constructed by using the global symmetries
of the higher-dimensional theory: as a result, a non-semisimple subgroup 
of these is gauged.
Supersymmetry breaking can be partial or total, depending on the
choice of the twist parameters. 
It is also possible to halve the number of surviving supersymmetries 
without reducing the number of independent breaking parameters 
by adding an orbifold projection (see for example \cite{vz,alten}).
In this way we obtain flat-gauged $\Ne6$, $\De4,5$, and $\Ne3$, $\De4$
no-scale models with three independent mass scales.

The plan of the paper is as follows. In the next section we review the structure
of $\Ne6$ supergravities in diverse dimensions, the field contents, 
the symmetries and earlier results. In section~\ref{sec:genred}, we describe
generalized reductions from $D$ to $D-1$ dimensions, discussing the general 
properties of the method. The details of the $\De5\to4$ and $\De6\to5$ cases
are presented in sections~\ref{sec:d5} and \ref{sec:d6} respectively, while 
orbifold is discussed in section~\ref{sec:orbi}. The last section 
contains our conclusions. 

\section{$\bf \Ne6$ in diverse dimensions}

$\Ne6$ supergravities did not receive as much attention as their
$\Ne8$ and $\Ne4$ sisters, since no string theory exists with
such an amount of supercharges in its unbroken phase. 
However, it has been shown \cite{n6d4str} that existing superstring theories
can be reduced to $\Ne6$ supergravities in four dimensions via asymmetric orbifolds.
Beyond four dimensions, $\Ne6$ supergravities exist only in $\De5$ and $\De6$.
The $\De5$ theory was first derived in \cite{cremmer}, and its $AdS_5$ gauged version 
was constructed in \cite{AdS5n6}. In $\De6$ two theories exist, the 
(6,0) and the (4,2). The former has neither graviton nor gravitino and
will not be considered. The latter, constructed in \cite{n6d6-1,n6d6-2},
is anomalous and has no lagrangian, as the (2,0) and the (4,0).
The anomaly can be cancelled only by adding two gravitino supermultiplets, thus
recovering the $\Ne8$ supergravity \cite{n6d6-2}. However, this does not represent a
problem since we are interested only in the effective $\De5$ theory obtained 
by dimensional reduction. 

We can summarize the field content of the three theories [$\De4$, $\De5$ and the (4,2) of $\De6$] 
as
%
\begin{eqnarray*}
\De4&:&[{\bf 1,6,15+1,20+6,15+\overline{15}}]_0 \\
&&\ [e_{\mu}^{\ \alpha},\psi_\mu^I,V_\mu^{IJ}+V_\mu,\chi^{IJK}+\chi^I,\phi^{IJ}+\bar\phi_{IJ}]_0 \\
\De5&:&[{\bf 1,6,14+1,14+6,14}]_0 \\
&&\ [e_M^{\ A},\psi_M^i,V_M^{ij}+V_M,\chi^{ijk}+\chi^i,\phi^{ij}]_0 \\
\De6&:&[{\bf 1,(4,1)+(1,2),(5,1)_2^- +(1,1)_2^+ +(4,2),(5,2)+(4,1),(5,1)}]_0 \\
&&\ [e_{\dot M}^{\ \dot A},\psi_{\dot M}^a+\psi_{\dot M}^{\rm a},V_{\dot M \dot N}^{- \ ab}+V_{\dot M \dot N}^+ 
+V_{\dot M}^{a,\rm a},\chi^{a b, \rm a}+\chi^{a},\phi^{ab}]_0 \,;
\end{eqnarray*}
%
we used the following short-hand notation for supermultiplets
\beq
[(J=2),(J=3/2),(J=1),(J=1/2),(J=0)]_{0/m} \,, \nn
\eeq
where the subscript $(_{0/m})$ indicates a massless or massive multiplet,
$(J=n)$ indicates, for each value of the spin $J=n$, the field representations 
of the corresponding automorphism groups~\footnote{For simplicity, in the rest of the paper
we will report only the number of states for each spin $J=n$ without making field representations
explicit.}, $U(6)$, $USp(6)$ and  $USp(4)\times USp(2)$ and finally the subscript $(_2)$ 
stands for a two-form.
Space-time indices $\mu,\nu...$, $M,N...$, $\dot M,\dot N...$ refer to $\De4,5,6$,
while the indices $I,J...$, $i,j...$, $a,b...$, $\rm a,b...$ refer to $SU(6)$, $USp(6)$, $USp(4)$ 
and $USp(2)$ respectively. Repeated indices belonging to the same group are antisymmetrized 
and, in the case of $USp(2N)$, they are also $\Omega$-traceless 
($\Omega$ being the $USp(2N)$ symplectic metric).
The 2-forms $V_{\dot M \dot N}^{(-/+)}$ are (anti) self-dual antisymmetric 
tensors in six dimensions.
These theories, besides supersymmetry and the invariance 
under general coordinate transformations, have a non-compact global symmetry,
called $U$-duality ($G$), whose maximal compact subgroup is represented by the $R$-symmetry group ($H$).
The manifolds $G/H$ define the following non-linear $\sigma$-models
\bea
&&\hspace{-0.2cm} {\cal M}_{\De4}=\frac{SO^*(12)}{U(6)} \,, \qquad {\cal M}_{\De5}=\frac{SU^*(6)}{USp(6)} \,, 
\nn \\ && 
{\cal M}_{\De6}=\frac{SU^*(4)\times SU^*(2)}{USp(4)\times USp(2)} \simeq  \frac{SO(5,1)}{SO(5)}\,,
\eea
for the scalar fields in each theory~\footnote{More details on groups $SU^*(2N)$, $SO^*(2N)$ 
and their algebras can be found, for example, in \cite{gilmore}.}.

\section{Generalized reduction} \label{sec:genred}
Generalized reduction from $D$ to $D-1$ dimensions corresponds to imposing twisted 
periodicity conditions along the compact dimension, namely
\beq
\Phi(x^\mu,y+2\pi r) = U \, \Phi(x^\mu,y) \,,
\eeq 
with $U=\exp[T]$ is a symmetry of the theory. To switch back to periodic fields, it is sufficient 
to rescale them with the factor $U(y)=\exp[y/(2\pi r) T]$.
If the twist is chosen to be in the $U$-duality compact subgroup, 
supersymmetry can be broken spontaneously without producing a cosmological constant 
(non-compact twists, in general, lead to theories with unstable scalar potentials and therefore will not be considered here).
The gauged group generated in the effective theories is a non-semisimple subgroup of the $U$-duality group,
in particular the semidirect product of a $U(1)\subset H$ (the Scherk--Schwarz twist $U$) and $n$ shift symmetries, corresponding to
the higher-dimensional gauge invariance along the compact direction. This can be easily deduced by
looking at the reduction of field strengths and kinetic terms. It is easy to verify, 
for instance, that field strengths with flat indices 
$V_{AB}=e_A^{\ M} e_B^{\ N}\partial_{[M} V_{N]}$ [$M=(\mu,y),A=(\alpha,\hat y)$] reduce as
\bea
V_{\alpha \beta} &\to& \rho^{-2\gamma} e_\alpha^\mu e_\beta^\nu\, U(y) \l ( B_{\mu\nu}+M B_{[\mu} A_{\nu]}+V_y A_{\mu\nu} \r )
= \rho^{-2\gamma} e_\alpha^\mu e_\beta^\nu\, U(y) \l ( {\hat B}_{\mu\nu}+V_y A_{\mu\nu} \r ) \,,\nn \\
V_{\alpha \hat y} &\to& \rho^{-\gamma-1} e_\alpha^\mu\, U(y) \l [\partial_\mu V_y - M \l (B_\mu +A_\mu V_y \r )\r ] 
=  \rho^{-\gamma-1} e_\alpha^\mu\, U(y) D_\mu V_y \,, \label{eq:albe}
\eea
where we have not reported internal indices explicitly,
\bea
M&\equiv&U(y)^{-1}\partial_y U(y) \,, \nn \\
B_{\mu}&\equiv&V_{\mu}-V_y A_{\mu} \,,  \nn \\
{\hat B}_{\mu\nu}&\equiv& B_{\mu\nu}+M \, B_{[\mu} A_{\nu]} \,, 
	\qquad B_{\mu\nu}\equiv\partial_{[\mu} B_{\nu]}\,, \qquad A_{\mu\nu}\equiv\partial_{[\mu} A_{\nu]}\,, 
\eea
$\rho=e_y^{\ \hat y}$ is the radion and $A_\mu = \rho^{-1} e_\mu^{\ \hat y}$ is the 
graviphoton~\footnote{In deriving eq.~(\ref{eq:albe}) we adopted the vielbein parametrization of \cite{ss},
i.e. the triangular gauge $$e_M^{\ A}=\l( (e_y^{\ \hat y})^\gamma\,e_\mu^{ \ \alpha},\, e_\mu^{\ \hat y} ,\, 0\,,\, e_y^{\ \hat y} \r)\,.$$}.
The $(_{\alpha\beta})$ term in eq.~(\ref{eq:albe})
produces field strengths that are covariant with respect to the group $U(1)\circledS {\cal T}^n$,
with structure constants given by the Scherk--Schwarz mass matrix $M$. 
The compact factor is gauged by the graviphoton coming from the higher-dimensional metric,
while the translation symmetries were gauged by the corresponding vectors. The term $(_{\alpha \hat y})$, 
on the other hand, produces the covariant kinetic term for the axions $V_y$, which can be
shifted away in the unitary gauge, giving mass to the vectors through the Higgs mechanism. 
As discussed in \cite{cs,fg28,vz} Chern--Simons terms, besides getting covariant, produce
extra contributions which are necessary for the invariance of the lower-dimensional theory.

It is straightforward to verify that the generalized reduction works in the same way 
for the fermionic sector.
In fact, the $(_{\alpha\beta})$ term of the flat derivative $e_A^{\ M} e_B^{\ N}\partial_{[M} \psi_{N]}$
makes the lower-dimensional gravitino kinetic term, covariant with respect to the gauged group, while the $(_{\alpha \hat y})$ 
one gauges the $\psi_y$ component. Analogously to $V_y$ with the vector fields $V_\mu$, $\psi_y$ is reabsorbed by
gravitinos, which get the mass $M$ through the super-Higgs effect.
Finally the scalar manifold vielbein ${\cal P}_\mu$ becomes covariant as well,
while the extra component produces a potential ${\rm tr}\,({\cal P}_y^2)$ that vanishes at its extremum.
  
The gauged supergravity obtained in this way is then a spontaneously broken one, with vanishing vacuum energy.
We now discuss more in detail the $\Ne6$ case.
\section{The $\De4$ effective theory} \label{sec:d5}
Although the four-dimensional $\Ne6$ $U$-duality group is $SO^*(12)$, when we consider 
the effective theory obtained by dimensional reduction from the $\De5$ one,
only part of this group will be a symmetry of the action. It is then useful 
to decompose the corresponding algebra with respect to the five-dimensional one [$\fsu^*(6)$]:
\beq
\fso^*(12)=\fusp(6)\,\oplus\, [\fsu^*(6) \ {\rm mod} \ \fusp(6)]\, \oplus\, \fso(1,1)
+({\bf 14}+{\bf 1})_\tau + ({\bf 14}+{\bf 1})' \,,
\eeq
on the right-hand side of the above equation, we have respectively:
 the $R$-symmetry inherited from the five-dimensional one,
the algebra of the $\De5$ scalar manifold, the dilatation $\fso(1,1)$  
associated with the radion $e_5^{\ \hat 5}$, and the $({\bf 14+1})_\tau$ shift symmetries
coming from the corresponding $\De5$ vector fields. 
The remaining $({\bf 14+1})'$ generators are the compact elements that would extend $USp(6)$
to $U(6)$, but now fall into the magnetic subgroup of the $Sp(32,{\bf R})$ duality group 
acting on the vector field strengths and their duals \cite{gz}. 
These generators can still be rotated into the 
electric subgroup, recovering the $U(6)$ symmetry of the action, 
by means of a duality transformation.
However, when the gauging will be turned on via the Scherk--Schwarz mechanism, 
the embedding showed above will be frozen, 
and the $U(6)$  and $USp(6)$ formulations will no longer be equivalent.

The scalar geometry can be decomposed as well:
\beq
\frac{SO^*(12)}{U(6)} = \l [ \frac{SU^*(6)}{USp(6)}\times SO(1,1)\r]\circledS ({\cal T}^{14+1}) \,,
\eeq
showing how $\De5$ scalar fields fit into the four dimensional coset manifold.

The rank-3 group $USp(6)$ can thus be used for the twist, 
producing a mass matrix $M=U^{-1}(y)\partial_y U(y)$ with three independent mass parameters $m_{1,2,3}$.

The spectrum is twice degenerate, as this mechanism cannot produce chirality. 
The mass for each field can be read directly from $M$ and depends only on the way the field transforms under $USp(6)$.
The mass spectrum, already deduced in 
\cite{sH}~\footnote{We take this opportunity to correct a typo in table 7 of \cite{sH}.},
 is summarized in Table~\ref{tab:spectrum-d4} and satisfies the mass formula 
 ${\rm str} {\cal M}^4={\rm str} {\cal M}^2=0$ \cite{massfor}.
\begin{table}[t]
\begin{center}
\begin{tabular}{|c|c|c|c|}
\hline
Field & Mass & \multicolumn{2}{c|}{No. of states} \\ 
& & $\Ne6$ & $\Ne3$ \\
\hline\hline 
$e_\mu^{\ \alpha}+e_\mu^{\ \hat 5}+e_5^{\ \hat 5}$ & $0$ & $ 1+1+1$ & $1+0+1$\\ \hline 
$\psi_\mu^i+\psi_5^i$ & $|m_{1,2,3}|$ & $ 6+6$  & $3+3$\\ \hline
$V_\mu^{ij}+V_5^{ij} $ & $|m_\ell \pm m_{\ell'}| \quad \ell>\ell' $&$12+12$ & $6+6$ \\ 
	& $ 0 $&$ 2+2$  & $0+2$ \\ \hline
$V_\mu+V_5$ & $0 $&$1+1$ & $0+1$ \\ \hline
$\chi^{ijk}$ & $|m_1 \pm m_2 \pm m_3| $&$ 8$ & $4$\\
 	& $|m_{1,2,3}| $&$ 6$ & $3$ \\ \hline
$\chi^i$ & $|m_{1,2,3}| $&$ 6$ & $3$\\ \hline
$\phi^{ij}$ & $|m_\ell \pm m_{\ell'}| \quad \ell>\ell' $&$12$ & $6$ \\ 
	& $ 0 $&$ 2$ & $2$	\\
\hline 
\end{tabular}
\caption{\label{tab:spectrum-d4} {\it Spectrum of $\De4$ supergravity 
obtained by generalized reduction of $\Ne6$, $\De5$ supergravity. 
The last two columns give the number of states for the reduced $\Ne6$ and $\Ne3$ theories respectively.}}
\end{center}
\end{table}
Depending on the number of mass parameters turned on, supersymmetry can be broken 
totally or partially to $\Ne4,2,0$. The gauged group can be $U(1)\circledS {\cal T}^{12,10,8,6}$,
depending on the number of massive vectors. 

The broken phases then appear to be:
\begin{eqnarray*} 
 &\Ne6\to4& \\&  [1,4,6,4,2]_0 + \{ 2\times[0,1,4,6,4]_m \} + 2\times[0,0,1,4,6]_0&  \\ \\ 
 &\Ne6\to2& \\& \hspace{-0.15cm} [1,2,1,0,0]_0 + 3\times[0,0,1,2,2]_0 + 2\times\{ 2\times[0,1,2,1,0]_m \} &\\
 	&+\,2\times\{ 2\times[0,0,1,2,1]_m \}+2\times\{ 2\times[0,0,0,1,2]_m \}   &  \\ \\
 &\Ne6\to0& \\& [1_0,6_m,12_m+4_0,20_m,12_m+6_0] & 
\end{eqnarray*}
where, as expected \cite{dual1,sH}, massive multiplets in curly brackets 
are $\frac12$BPS {\em short} multiplets. 

These theories thus represent the $\Ne6$ no-scale models in four dimensions.
\subsection{Orbifold projection} \label{sec:orbi}
We can also obtain chiral theories by compactifying on the orbifold $S^1/\Zd$. 
The orbifold reduction can be implemented by assigning $\Zd$ parities to fields
\beq
\Phi(-y)=\Zd \Phi(y) \,,
\eeq
 in such a way that the theory remains invariant under the transformation 
$y\to-y$, and eliminating odd-parity fields that have no zero modes.
This projection produces chirality by halving the number of gravitinos 
(and thus the number of supersymmetries). 
The naive orbifold reduction leads to an $\Ne3$ supergravity coupled
to three vector multiplets, whose scalar geometry is
\beq
\frac{SU(3,3)}{U(3)\times SU(3)} \,.
\eeq
A Scherk--Schwarz twist can be turned on by using a Cartan 
subalgebra of $USp(6)$, which anticommutes with $\Zd$, along the lines of \cite{vz}
for the $\Ne4$ case. The resulting theory is a matter-coupled flat $\Ne3$ supergravity,
spontaneously broken to $\Ne2,1,0$, with three independent mass parameters. 
The spectrum is reported in Table~\ref{tab:spectrum-d4} and is equal to the $\Ne6$ one
without the four massless vectors and with the degeneracy of the massive states halved.
Because the graviphoton $A_\mu$ is always odd under the $\Zd$ parity, the gauged group 
is an abelian ${\cal T}^n$ ($n$ being the number of massive vectors).
The field content for each theory is then:
\begin{eqnarray*} 
&\Ne3\to2& \\& [1,2,1,0,0]_0 + [0,1,4,6,4]_m + [0,0,1,2,2]_0+[0,0,0,2,4]_0 & \\ \\
&\Ne3\to1& \\& [1,1,0,0,0]_0 + 2\times[0,1,2,1,0]_m +2\times[0,0,1,2,1]_m & \\
	& +\, 2\times[0,0,0,1,2]_m  + 3\times[0,0,0,1,2]_0 &  \\ \\
&\Ne3\to0& \\& [1_0,3_m,6_m,10_m,6_m+6_0] &
\end{eqnarray*}
where the massive fields are now arranged into {\em long} non-BPS multiplets \cite{dual1,sH}.

We can recognize in this theory the one constructed by 
Tsokur and Zinovev by dualization \cite{tz3}, the derivation given here 
shows in a simple way its five-dimensional origin, and explains why
its spectrum still satisfies the $\Ne6$ supersymmetry 
constraint ${\rm str}{\cal M}^4={\rm str}{\cal M}^2=0$.

This theory can also be obtained from a $T^6/\Zd$ orientifold reduction 
of Type IIB supergravity with fluxes turned on \cite{fp,fluxes1}.
Then we found another {\em low-energy duality} between flux and Scherk--Schwarz 
compactifications, analogous to the one found in \cite{fluxes3} for the $\Ne4$ case.
\section{The $\De5$ effective theory} \label{sec:d6}
Finally we discuss the $\De6\to5$ case.
As before, it is useful to look at the decomposition of the $\De5$
$U$-duality algebra $\fsu^*(6)$ in terms of the $\De6$ one [$\fsu^*(4)\oplus \fsu^*(2)$]:
\beq
\fsu^*(6)=\fusp(4)\, \oplus\,\fusp(2)\,\oplus\,[\fsu^*(4) \ {\rm mod} \ \fusp(4) ]\, \oplus \ \fso(1,1) 
+ ({\bf 4},{\bf 2})_\tau + ({\bf 4},{\bf 2})' \,,
\eeq
where it is simple to identify respectively the $\De6$ $R$-symmetry, the coset algebra of the higher-dimensional
scalar manifold, the dilatation associated to the modulus $e_6^{\ \hat 6}$, the eight shift generators 
in the $({\bf 4},{\bf 2})$ of $USp(4)\times USp(2)$ associated to the corresponding 
$\De6$ vector fields, and the compact generators of  $USp(6)/[USp(4)\times USp(2)]$.
The scalars of the $\De6$ coset manifold, together with those coming from the compactification procedure, 
are thus embedded into the $\De5$ one as follows:
\beq
\frac{SU^*(6)}{USp(6)} = \l [ \frac{SU^*(4)}{USp(4)} \times SO(1,1) \r]\circledS {\cal T}^{(4,2)} \,.
\eeq
Three independent 
parameters are still available for the twist [$m_{1,2}$ from $USp(4)$ and $m_3$ from $USp(2)$].
The spectrum can easily be derived by looking at how fields transform under the $R$-symmetry group,
and is reported in Table~\ref{tab:spectrum-d5}.
\begin{table}[t]
\begin{center}
\begin{tabular}{|c|c|c|}
\hline
Field & Mass & No. of states \\
\hline\hline
$e_M^{\ A}+e_M^{\ \hat 6}+e_6^{\ \hat 6}$ & $0$ & $ 1+1+1$ \\ \hline 
$\psi_M^a+\psi_6^a$ & $|m_{1,2}|$ & $ 4+4$ \\  \hline
$\psi_M^{\rm a}+\psi_6^{\rm a}$ & $|m_3|$ & $ 2+2$ \\ \hline
$V_{MN}^{ab}$ & $|m_1 \pm m_2|$ & $2_2^{\bf c}$ \\
	& $0$ & $1$ \\ \hline
$V_{MN}$ & $0$ & $1$ \\ \hline
$V_M^{a,\rm a}+V_6^{a, \rm a} $ & $|m_{1,2} \pm m_3|  $&$8+8$ \\ \hline
$\chi^{ab,\rm a}$ & $|m_1 \pm m_2 \pm m_3| $&$ 8$ \\
 	& $|m_3| $&$ 2$ \\ \hline
$\chi^a$ & $|m_{1,2}| $&$ 4$ \\ \hline
$\phi^{ab}$ & $|m_1 \pm m_2|  $&$4$ \\ 
	& $ 0 $&$ 1$ \\ \hline
\end{tabular}
\caption{\label{tab:spectrum-d5} {\it Spectrum of the $\Ne6$, $\De5$ reduced supergravity}}
\end{center}
\end{table}
Note that four anti self-dual forms complexify into two massive complex
antisymmetric tensors fields in $\De5$ ($2_2^{\bf c}$), while the remaining 
two can be dualized to vectors.
As in $\De4$, we can have partial or total supersymmetry breaking to $\Ne4,2,0$. 
Note that $\Ne4$ can be obtained in two different ways:
by turning on $m_3$ we get a $USp(4)$ invariant action, while turning on either $m_1$ or $m_2$ 
we have only $USp(2)\times USp(2)$. 
The gauged groups are $U(1) \circledS{\cal T}^{8,6,4}$, depending on the number of vectors acquiring a mass.
The broken phases for $\Ne6\to4,2,0$ in $\De5$ then are
\begin{eqnarray*}
 &\Ne6\to4 \quad USp(4)  & \\ &[1,4,6,4,1]_0 + \{ 2\times[0,1,4,5,0]_m \} + [0,0,1,4,5]_0  &  \\ \\
 &\Ne6\to4 \quad USp(2)\times USp(2)& \\ &[1,4,6,4,1]_0 + \{ 2\times[0,1,2+1_2^{\bf c},5,2]_m \} + [0,0,1,4,5]_0  &  \\  \\
 &\Ne6\to2& \\& [1,2,1,0,0]_0+2\times[0,0,1,2,1]_0   &\\
 	&\,+ 2\times\{ 2\times[0,1,2,1,0]_m \}+\{ 2\times[0,0,1_2^{\bf c},4,2]_m \}&  \\ \\
 &\Ne6\to0& \\&[1_0,6_m,(2_2^{\bf c})_m + 8_m+3_0,14_m,4_m+2_0] & 
\end{eqnarray*}
In this case anomalies will not allow us to upgrade the result to six dimensions.
We simply used the $\De6$ theory together with generalized dimensional reduction
as a tool to derive the $\De5$, $\Ne6$ no-scale model, analogously to what had been done in \cite{vz} for the 
(4,0) $\De6$ theory.
\section{Conclusions}
We constructed $\Ne6$ no-scale models in $\De4,5$ through
generalized dimensional reduction, and discussed the gauging, 
the spectrum and the scalar geometries. The theories obtained 
in this way belong to a class of gauged supergravities not considered
by earlier classifications. Moreover the theories
presented in this paper, together with the one 
constructed in \cite{AdS5n6}, are the only known 
$\Ne6$ gauged supergravities. We also rederived the 
$\Ne3$, $\De4$ theory of \cite{tz3}, giving it a 
five-dimensional interpretation.
\section*{Acknowledgements}
We are grateful to S.~Ferrara and F.~Zwirner for helpful discussions and comments on the manuscript.
\end{document}